\documentclass{optica-article}

\journal{opticajournal} % for journals or Optica Open

\articletype{Research Article}

\usepackage{lineno}
\usepackage{graphicx}
\usepackage{mathtools} % already loads amsmath

\usepackage{amssymb}

\usepackage{
	bm,
	titlesec,
	braket,
	dutchcal,
	xcolor,
	upgreek,
	soul
}
\usepackage{bm}

\newcommand{\bmsection}[1]{\section*{#1}}

\newcommand{\sket}[1]{\ket{#1}_s}

\newcommand{\sbra}[1]{{\raisebox{-0.7ex}{${\scriptstyle s}$}}\!\bra{#1}}

\begin{document}

\title{Coherent temporal filtering of multimode parametric down-conversion using a quantum pulse gate}

\author{Abhinandan Bhattacharjee$^{*}$,  Patrick Folge, Laura Serino\authormark{\textdagger}, Sebastian Lengeling, Benjamin Brecht, and Christine Silberhorn}

\address{Integrated Quantum Optics Group, Institute for Photonic Quantum Systems (PhoQS), Paderborn University, Warburger Straße 100, 33098 Paderborn, Germany}

\address{\authormark{\textdagger} Present address: School of Mathematics and Physics, University of Queensland, Brisbane, Australia}

\email{\authormark{*}abhib@mail.uni-paderborn.de} %% email address is required; see note below about the corresponding author designation

% use {asbstract*} to suppress the copyright line. Copyright information will be added in production

\begin{abstract*} 
Spectrally pure and indistinguishable single photons are essential for quantum network platforms, where high-visibility interference underpins many quantum information protocols. However, most practical single-photon sources emit spectrally multimode states with reduced purity. Conventional spectral intensity filtering can partially improve purity but cannot select a well-defined temporal mode (TM). Here, we demonstrate coherent temporal filtering of a multimode parametric down-conversion (PDC) source using a quantum pulse gate (QPG) and benchmark its performance against conventional intensity filtering. The generated PDC photons exhibit strong spectral correlations, rendering extraction of pure heralded photons from the pair a challenge. We demonstrate that QPG filtering consistently generates heralded photons with purities above 0.90 regardless of the filter shape. Contrariwise, using spectral intensity filters yields mixed photons. Photon purities are probed with chronocyclic Q-function tomography. Furthermore, we demonstrate the versatility of QPG filtering by extracting structured TMs, including superposition of picosecond time bins. These results establish the QPG as a practical coherent filtering tool for quantum network applications. 

\end{abstract*}

\section{Introduction}
Pure single-photon states have emerged as a fundamental resource for optical quantum network platforms. They are the building blocks of a wide range of quantum information science applications, including entanglement swapping \cite{zukowski1993prl, pan1998prl}, quantum teleportation \cite{ma2012natphy}, boson sampling \cite{tillmann2013nat, spring2013science}, and photonic quantum walks \cite{sansoni2012prl, peruzzo2010science}. The performance of such platforms relies on high-visibility interference between pure and indistinguishable single photons. Consequently, the development of sources emitting photons in well-defined spatial, temporal, and polarization modes is essential. In this context, fiber- and waveguide-based sources naturally produce photons in a single spatial and polarization mode, ensuring high mode purity in these degrees of freedom \cite{harder2016prl, spring2017optica, eigner2020optexp}. Achieving high spectral purity therefore remains a central requirement for scalable photonic quantum technologies.

Spectrally pure single-photon states occupy a well-defined temporal mode (TM) \cite{brechtprx2015}. In such a mode, stable phase relations exist between the spectral components, resulting in perfect mutual coherence. Single-photon sources, including heralded parametric down-conversion (SPDC) \cite{de2003pra,kaltenbaek2006prl,PhysRevA.81.021801,vergyris2016scirpt} and spontaneous four-wave mixing (SFWM) \cite{harada2011injp,koefoed2017pra} sources and solid-state emitters \cite{aharonovich2016natphot, senellart2017high}, typically produce photons in incoherent mixtures of multiple TMs with fluctuating phase relations between their spectral components. Consequently, the mutual coherence between spectral components is reduced, leading to lower spectral purity. These mixed multimode states therefore limit the performance of quantum network platforms that rely on high-visibility interference. 

A common approach to improve spectral purity is narrowband spectral intensity filtering \cite{pan1998prl,silverstone2014natphot,thomas2026praapp,suarez2020josab}. However, such filtering acts only on spectral amplitudes and does not control the relative phase between spectral components, and therefore constitutes an incoherent operation. Consequently, it cannot distinguish different TMs or reject spectrally overlapping noise within the transmission bandwidth. The resulting trade-off between noise rejection and single-mode purity has been identified as a fundamental constraint in practical quantum network deployments \cite{thomas2026praapp}. Perfect spectral purity can only be approached in the limit of an infinitesimally narrow transmission bandwidth \cite{christ2013njp, christ2014pra, meyer2017pra}, which is impractical for most applications. Moreover, intensity filtering cannot programmably select or prepare temporal modes with arbitrary bandwidths and shapes. Overcoming these limitations requires mode-selective filtering, that is, coherent filtering, whose advantages over incoherent implementations have been analyzed theoretically in Ref.~\cite{raymer2020oe}. This approach can extract well-defined TMs and enable the preparation of pure single-photon states from multimode sources. 

Coherent filtering is a well-developed technique in the radio-frequency domain \cite{hlawatsch2002time,matz2018linear}, where carrier frequencies lie in the MHz-GHz range. In this regime, filters can readily operate on microsecond to nanosecond timescales. Optical fields, however, oscillate at petahertz frequencies and require filter responses on the femtosecond timescale, which is difficult to realize experimentally. This makes the implementation of coherent filtering for broadband optical photons particularly challenging.  The quantum pulse gate (QPG) \cite{brecht2014pra, ansari2018prl}, a dispersion-engineered sum-frequency conversion process that enables high-fidelity TM projections of pulsed single photons, overcomes this challenge. The QPG has been employed for TF tomography \cite{ansari2018prl, gil2021optical, serino2025self}, single-photon pulse characterization \cite{bhattacharjee2025optexp, bhattacharjee2025frequency}, quantum optical coherence tomography \cite{namekata2023scirpt}, high-resolution temporal discrimination \cite{donohue2018prl, ansari2021prxquant} and TM-selective detection in optical communication \cite{jarzyna2023ieee}. Notably, coherent filtering of narrowband photons has also been demonstrated using atomic quantum-memory platforms \cite{gao2019prl, thomas2024sciadv, zhang2025arxiv}.

\begin{figure*}[t!]
	\centering
	\includegraphics[width=\textwidth]{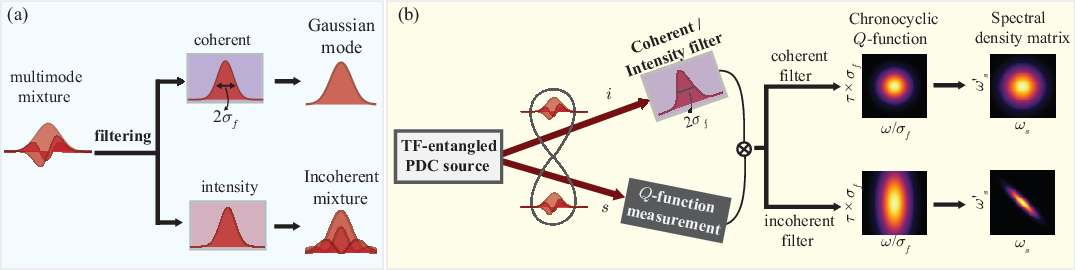}
	\caption{(a) Conceptual distinction between coherent filtering and intensity filtering. (b) Implementation of coherent and intensity filtering on a time–frequency entangled parametric down-conversion (PDC) source. Filtering is applied to the heralding (idler) photon ($i$). The effect of filtering is characterized through the chronocyclic $Q$-function of the signal photon ($s$), from which the corresponding spectral density matrix is reconstructed.}\label{fig1}
\end{figure*} 
In this article, we demonstrate remote temporal-mode preparation from a spectrally entangled multimode type-0 PDC source using coherent filtering with a QPG, whereas conventional intensity filtering always produces a multimode mixture. The PDC source generates time-frequency entangled photon pairs whose spectral correlation bandwidth is significantly narrower than the marginal bandwidths of individual photons. Non-frequency-resolving detection of one photon yields a spectrally mixed heralded photon. We apply either a QPG or an incoherent (intensity) filter to one photon of the pair and detect it after filtering. This measurement remotely projects the heralded partner photon onto either a well-defined TM or a multimode mixture, depending on the filtering operation. 

In our first experiment, we employ Gaussian filters with bandwidths ranging from the PDC correlation bandwidth to significantly larger values and verify experimentally how coherent versus incoherent filtering impacts the purities of heralded single-photon states. Our results show that QPG filtering yields Fourier-limited Gaussian modes across the entire bandwidth range with purity above 0.9. In contrast, incoherent filtering produces multimode mixtures with progressively reduced purity as the bandwidth increases. In our second step, we demonstrate how the programmability of the QPG can be harnessed to extract Hermite-Gaussian (HG) TMs and picosecond time-bin superpositions from the multimode PDC source for programmable remote preparation of heralded single-photon states.

\section{Theory}
\subsection{Concept}
Figure~\ref{fig1}(a) illustrates the conceptual difference between coherent and incoherent filtering of TMs. A coherent filter programmed with a Gaussian mode profile transmits a single Gaussian TM from the multimode input, while rejecting all other TMs. In contrast, an intensity filter with a Gaussian transmission window transmits the spectral intensities from all TMs while erasing the phase information, leading to an incoherent mixture.

Figure~\ref{fig1}(b) conceptually illustrates coherent and incoherent filtering applied to a multimode PDC source that produces TF-entangled signal $(s)$ and idler $(i)$ photons. The TF entanglement produces strong correlations between the TMs of the PDC photons. We apply either coherent or incoherent filtering to the idler photon and use it for heralding. This remotely shapes the TM structure of the signal photon \cite{ansari2020optexp}. The heralded photon is characterized using chronocyclic $Q$-function measurements \cite{bhattacharjee2025optexp}. Coherent and incoherent filtering produce distinctly modified the chronocyclic $Q$-functions of the heralded photon, although only coherent filtering enables controlled shaping. 

\subsection{Coherent and incoherent temporal-mode filtering}
The joint state of the photon pair can be written as \cite{grice1997pra} %
\begin{equation}
	\ket{\psi}  = \int F(\omega_s,\omega_i) \ket{\omega_s}_s\ket{\omega_i}_i d\omega_sd\omega_i,\label{psi}
\end{equation}
where $F(\omega_s,\omega_i)$ is the joint spectral amplitude (JSA), and $\omega_s$ and $\omega_i$ are the angular frequencies of signal and idler photons, respectively. The JSA fully describes the TF properties of the joint state. The simulated JSA is shown in Supplementary Material Fig.~S1

In this scheme, we employ either coherent or incoherent filtering on the idler photon and subsequently detect it. The detection of the filtered idler photon heralds the presence of the partner signal photon. The conditional state of the heralded signal photon is described by a density matrix whose properties depend on the type of filtering applied to the idler. It is expressed as 
\begin{equation}
	\hat{\rho}_{s} = \iint \Gamma(\omega_s,\omega'_s) \sket{\omega_s}\sbra{\omega'_s}d\omega_sd\omega'_s,\label{DM}
\end{equation}
where $\Gamma(\omega_s,\omega'_s)$ is the spectral density matrix of the heralded signal photon. It describes the coherence between different frequency components of the heralded photon and therefore determines whether the photon occupies a single temporal mode or a mixture of modes.

A coherent filter selects a specific TM of the idler photon and is described by the complex spectral amplitude $G(\omega_i)$. Due to the strong spectral correlations, heralding the idler photon after filtering remotely prepares the signal photon in a well defined TM. Mathematically, this operation is equivalent to projecting the JSA $F(\omega_s,\omega_i)$ onto the filter mode $G(\omega_i)$ of the idler photon. The spectral density matrix of the signal photon takes the form
\begin{equation}
	\begin{aligned}
		\Gamma(\omega_s,\omega_s') &=
		\iint d\omega_i\, d\omega_i'\,
		F(\omega_s,\omega_i)\,
		F^*(\omega_s',\omega_i')\,
		G(\omega_i)\,G^*(\omega_i') \\
		&= C(\omega_s)\,C^*(\omega_s').
	\end{aligned}
	\label{DM2}
\end{equation}
Here, the spectral density matrix $\Gamma(\omega_s,\omega'_s)$ is factorized as the product of complex spectral amplitudes $C(\omega_s)C^*(\omega'_s)$. Such a factorized form indicates that the heralded signal photon occupies a single TM with complex spectral amplitude $C(\omega_s)$. This behavior is reflected in the circular profile of the spectral density matrix corresponding to coherent filtering in Fig.~\ref{fig1}(b). The mathematical derivation of this result and the corresponding simulated density matrices are presented in Supplementary Material Sec.~2 A.

On the other hand, an incoherent filter transmits individual frequency components of the idler photon without fixed phase relations. Each frequency component $\omega_i$ is transmitted a with probability determined by the transmission spectrum $|G(\omega_i)|^2$. A transmitted idler frequency remotely projects the signal photon onto a corresponding TM. Because these frequencies are transmitted independently, heralding performs an incoherent average over these projections. The spectral density matrix of the heralded signal photon therefore becomes
\begin{equation}
	\Gamma(\omega_s,\omega'_s) = \int F(\omega_s,\omega_i)F^*(\omega'_s,\omega_i)|G(\omega_i)|^2 d\omega_i.\label{DM4}
\end{equation}
This expression, in general, cannot be factorized into the form $C(\omega_s)C^*(\omega'_s)$. The heralded signal photon corresponds to a statistical mixture of TMs rather than a single well-defined TM. The spectral density matrix shown in Fig.~\ref{fig1}(b) reflects this behavior: the diagonal width, representing the spectral bandwidth, is significantly larger than the off-diagonal width, which represents the spectral coherence. The corresponding derivation and simulated density matrices are presented in Supplementary Material Sec.~2 B.

\subsection{Chronocyclic $Q$-function}\label{Qfun-sec}
In the experimental implementation, we characterize the TF properties of the heralded signal photon through chronocyclic $Q$-function measurements \cite{bhattacharjee2025optexp}. It is defined as the projection of the input TM onto a coherent state in TF phase space, represented by a Fourier-limited Gaussian mode $\mathcal{E}_c(\omega_s,\omega,\tau;\sigma_c)$ of bandwidth $\sigma_f$ with spectral shift $\omega$ and time delay $\tau$. To represent the TF phase space in dimensionless coordinates, where the coherent state has equal widths along both axes, we apply the rescaling
\begin{equation*}
	\Omega = \frac{\omega}{\sigma_f}, \quad T = \tau \sigma_f, \quad \Omega_s = \frac{\omega_s}{\sigma_f}.
\end{equation*}
In the rescaled TF phase space $(\Omega,T)$, the chronocyclic $Q$-function can be expressed in terms of the spectral density matrix as 
\begin{equation}
	Q(\Omega,T) = \iint \Gamma(\Omega_s,\Omega'_s)\mathcal{E}^*_c(\Omega_s,\Omega,T)\mathcal{E}_c(\Omega'_s,\Omega,T)d\Omega_sd\Omega'_s.\label{Qfun2}
\end{equation}
The shape of the chronocyclic $Q$-function directly reflects the TM structure of the photon. A single Fourier-limited TM yields a symmetric distribution in TF phase space, whereas multimode mixtures lead to asymmetric and broadened distributions with increased marginal bandwidth product \cite{PhysRevA.75.023810}. 

The degree of multimode character is quantified by the spectral purity $P$ (see Supplementary Material Sec.~2 C), where $P=1$ corresponds to a single TM and $P<1$ indicates a multimode mixture. For Gaussian-shaped coherent and incoherent filters, the analytical relation between the bandwidth product of the $Q$-function marginals and the spectral purity is derived in Supplementary Material Sec.~4. This relation shows that an increasing bandwidth product corresponds to decreasing spectral purity.

\begin{figure*}[t!]
	\centering
	\includegraphics[width=\textwidth]{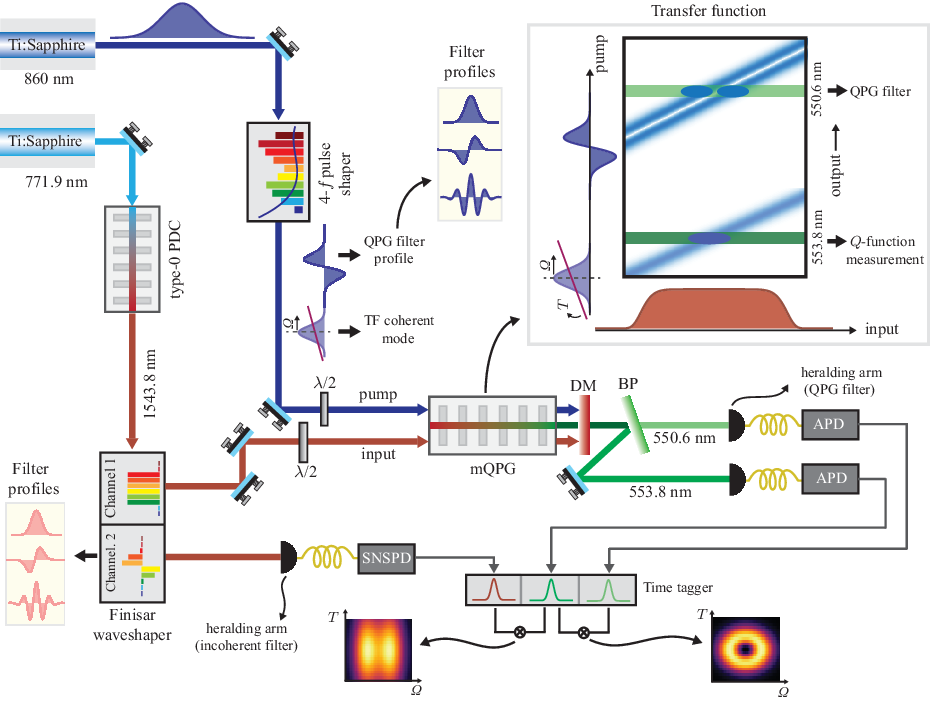}
	\caption{Schematic of the experimental setup. mQPG: multi-output quantum pulse gate, BP: bandpass, DM: dichroic mirror, APD: avalanche photodiode, SNSPD: superconducting nanowire single-photon detector}\label{fig2}
\end{figure*} 
\begin{figure*}[t!]
	\centering
	\includegraphics[width=\textwidth]{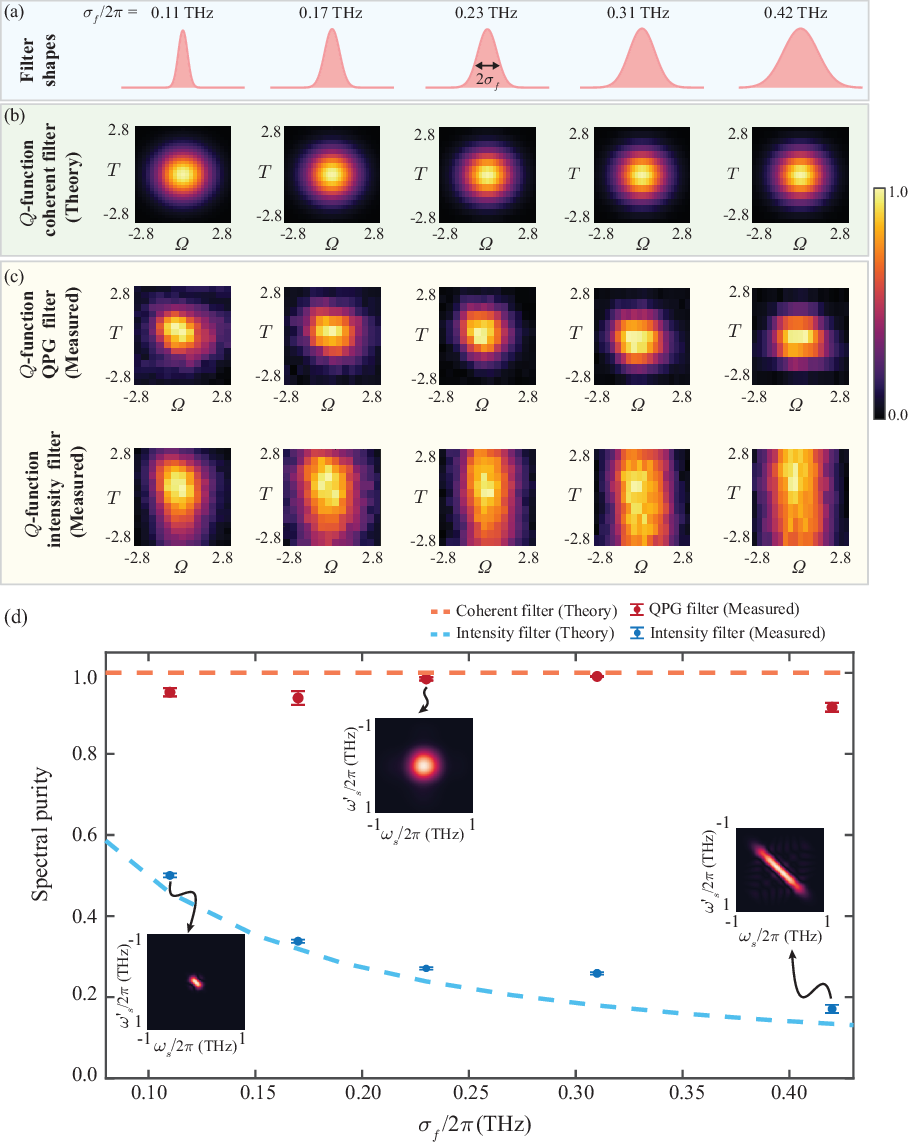}
	\caption{(a) Filter shapes with different spectral bandwidths. (b) Theoretical chronocyclic $Q$-functions corresponding to coherent filtering. (c) Measured chronocyclic $Q$-functions obtained using QPG filtering and intensity filtering. (d) Spectral purity as a function of filter bandwidth. The insets show the corresponding reconstructed spectral density matrices.}\label{fig3}
\end{figure*} 

\section{Experimental setup}\label{setup}
Figure~\ref{fig2} shows a schematic of the experimental setup. A Ti:Sapphire oscillator with a repetition rate of 76 MHz, a central wavelength of 771.9 nm, and a spectral bandwidth of 0.1 nm pumps a waveguide-integrated type-0 PDC source, which generates TF entangled photon pairs. The PDC source is a 1 cm long periodically poled titanium-in-diffused LiNbO$_3$ waveguide operated at a temperature of 433 K. In the following, we first describe the implementation of coherent filtering using a QPG, followed by the implementation of incoherent filtering.

For coherent filtering, the PDC photons are first spectrally filtered using a 20-nm-wide rectangular bandpass filter implemented with a Finisar waveshaper. They are then routed to a multi-output quantum pulse gate (mQPG) \cite{serino2023prx}. The mQPG comprises two up-converted output channels: one used for coherent filtering and the other for chronocyclic $Q$-function characterization. The two output channels are centered at 553.8 nm and 550.6 nm. The mQPG is a 4 cm long periodically poled titanium-in-diffused LiNbO$_3$ waveguide with a poling period of 4.32 µm, operated at 442 K. We use a second Ti:Sapphire oscillator with a central wavelength of 860 nm and a spectral bandwidth of 10 nm to pump the mQPG. This laser is repetition-rate-locked to the PDC pump laser. The pump pulses are shaped in amplitude and phase using a home-built 4-$f$ pulse shaper such that distinct spectral regions of the shaped pump independently address the two up-conversion channels of the mQPG (see the transfer-function illustration in Fig.~\ref{fig2}). The mQPG channel centered at 550.6 nm is used for coherent filtering by shaping the corresponding pump spectral region to a specific TM (a first-order Hermite Gaussian mode, as shown in Fig.~\ref{fig2}). We detect the up-converted output using an avalanche photodiode (APD), which serves as the heralding detector. The other mQPG channel, centered at 553.8 nm, is used for characterizing the chronocyclic $Q$-function of the heralded signal photon by scanning $(\omega, \tau)$ and detecting the up-converted output using a second APD. We use a dichroic mirror (DM) to suppress the residual pump and input from up-converted output and a 1 nm width bandpass filter is used to separate the two output channels. Coincidence detection between the two APDs yields the heralded chronocyclic $Q$-function for QPG-based coherent filtering.

For intensity filtering, the PDC output is instead routed to a multi-channel Finisar Waveshaper. The same spectral region is split into two separate channels. In one channel, we apply programmable filter shapes to implement incoherent filtering. The transmitted photons are detected using a superconducting nanowire single-photon detector (SNSPD), which serves as the heralding arm. In the other channel, a 20 nm wide rectangular bandpass filter is applied and the filtered photons are sent to the mQPG characterization channel (centered at 553.8 nm) for chronocyclic $Q$-function measurements, ensuring identical $Q$-function measurement settings across both filtering schemes. Coincidence detection between the mQPG output and the SNSPD yields the heralded $Q$-function corresponding to incoherent filtering (see Fig.~\ref{fig2}).

Each chronocyclic $Q$-function is measured using a $19\times19$ scan of the $(\omega,\tau)$ phase space, corresponding to 361 measurement points. The total acquisition time for each $Q$-function is approximately 6-8 h.

\begin{figure*}[t!]
	\centering
	\includegraphics[width=\textwidth]{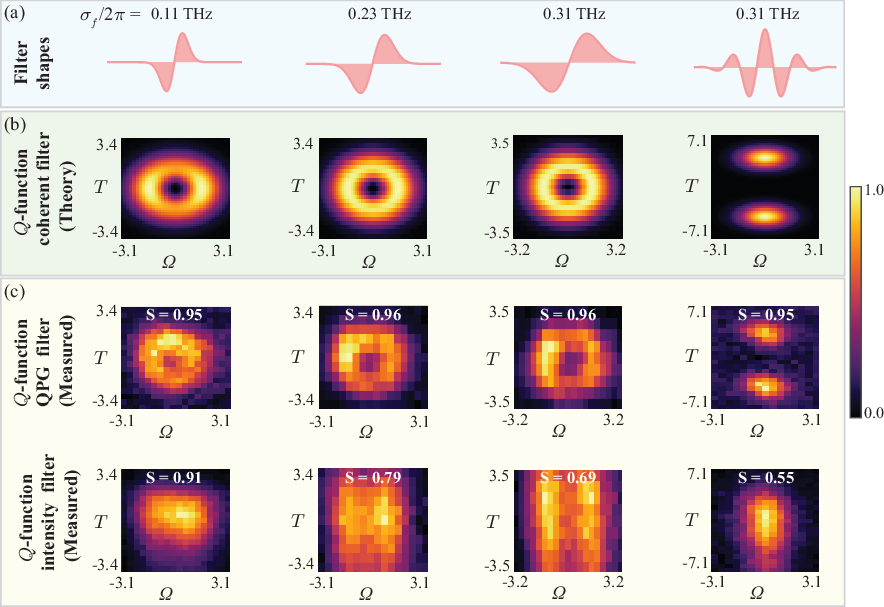}
	\caption{(a) Illustrates different filter shapes. (b) Theoretical chronocyclic $Q$-functions for a coherent filter. (c) Measured chronocyclic $Q$-functions corresponding to QPG filter and intensity filter. $S$ represents the similarity between theoretical and measured $Q$-functions.}\label{fig4}
\end{figure*}

\section{Results}\label{result}
We first implement Gaussian filters with bandwidths ranging from the PDC correlation bandwidth to significantly larger values (0.11-0.42 THz), as illustrated in Fig.~\ref{fig3}(a). Figure~\ref{fig3}(b) shows the expected chronocyclic $Q$-functions corresponding to coherent filtering for these shapes. Here, we rescale the phase space axes by $\sigma_f$, that is, $\Omega=\omega/\sigma_f$ and $T=\tau \times \sigma_f$. All the chronocyclic $Q$-functions become symmetric two-dimensional Gaussian distributions. The bandwidth product of their marginal distributions satisfies $\Delta \Omega \Delta T=1$, which is the hallmark of a Fourier-limited Gaussian TM. The spectral purity of each such TM is unity, as shown by Eq.~(S16) in the Supplementary Material. 

Figure~\ref{fig3}(c) shows the experimentally measured $Q$-functions for both QPG filtering and incoherent filtering. For QPG filtering, the measured chronocyclic $Q$-functions are in good agreement with theoretical predictions across the entire bandwidth range and retain the symmetric Gaussian shape expected for a single Gaussian TM. The marginal bandwidth products extracted from the experimental data are close to unity across the entire bandwidth range (see Supplementary Material Fig.~S6). These results indicate that QPG filtering extracts a Fourier-limited Gaussian TM. On the other hand, the measured chronocyclic $Q$-functions corresponding to incoherent filtering significantly differ from the coherent filtering predictions. They lose the symmetric Gaussian shape and become increasingly asymmetric and broaden along the  $T$-axis as the filter bandwidth increases. Consequently, their marginal bandwidth product, $\Delta \Omega \Delta T$, increases with filter bandwidth, indicating that the filtered state is no longer a single Fourier-limited Gaussian TM. Such broadening can, in general, arise either from a quadratic spectral phase or from a multimode mixture of TMs. A quadratic spectral phase introduces a tilt in the chronocyclic $Q$-function. However, no such tilt is observed in the measured $Q$-function. Therefore, the observed broadening is attributed to a multimode mixture whose contribution increases with filter bandwidth, indicating a decrease in the spectral purity of the filtered state, as derived in Supplementary Material Sec.~4. For clarity, the maximum of each $Q$-function is scaled to 1 in Fig.~\ref{fig3}.

In the context of quantum state tomography, it is well established that a quantum state is usually reconstructed from the phase space distributions \cite{d1994pra, landon2018prl}. Analogously, we reconstruct the spectral part of the density matrix $\Gamma(\omega_s,\omega'_s)$ from the measured chronocyclic $Q$-function data using a maximum-likelihood estimation procedure (See Supplementary Material Sec.~6) and evaluate the corresponding spectral purity $P$. The uncertainties in the reconstructed spectral purity are estimated using a Monte Carlo analysis based on Poissonian count statistics of the measured chronocyclic $Q$-functions. For each measured $Q$-function, 2000-5000 statistically independent realizations are generated, and the standard deviation of the reconstructed purity distribution is taken as the uncertainty. Figure~\ref{fig3}(d) shows the reconstructed spectral purity as a function of filter bandwidth for QPG and incoherent filtering. The dashed lines and dot markers represent the theoretical predictions and the reconstructed spectral purities, respectively. We find that the purity obtained with QPG filtering remains above 0.9 over the full investigated bandwidth range. In contrast, the purity associated with incoherent filtering decreases as the filter bandwidth increases. The insets display the representative reconstructed spectral part of the density matrices for the selected filter bandwidths. The complete set of reconstructed spectral part of the density matrices is presented in Supplementary Material Fig.~S7. For QPG filtering, the reconstructed density matrix has a symmetric two-dimensional Gaussian profile, representing a single Gaussian TM. The density matrices corresponding to incoherent filtering become increasingly diagonal, with the coherence width (off-diagonal width) becoming progressively smaller than the spectral bandwidth (diagonal width). This reduction in coherence width reflects an increasing multimode mixture contribution rather than a single TM. These results quantitatively demonstrate that QPG filtering extracts a single Gaussian TM over a wide bandwidth range from a multimode PDC source. These results clearly illustrate the fundamental distinction between coherent mode-selective filtering and conventional spectral intensity filtering. 

We next demonstrate the programmability of QPG filtering by extracting structured TMs beyond a Gaussian profile. For this purpose, we apply structured filter shapes: first-order Hermite-Gaussian (HG) modes with different bandwidths and  a second-order cosine-kernel (CK) function, as illustrated in Fig.~\ref{fig4}(a). For clarity, the second-order CK function represents an equal superposition of two picosecond time bins. Figure~\ref{fig4}(b) shows the theoretically expected chronocyclic $Q$-functions for coherent filtering corresponding to these filter shapes. In particular, for the CK mode, the predicted $Q$-function exhibits two well-separated distributions along the $T$-axis, which reflects the coherent superposition of two time-bins. Figure~\ref{fig4}(c) shows the experimentally measured $Q$-functions corresponding to QPG and incoherent filtering. The experimentally measured $Q$-functions exhibit close agreement with the theoretical profiles when QPG filtering is employed. For quantitative comparison, we evaluate the similarity $S$ between the theoretical and measured $Q$-functions, where $S=1$ implies perfect agreement and $S=0$ indicates no agreement. We find that $S \geq 0.95$ for all $Q$-functions corresponding to QPG filtering. For incoherent filtering, each measured $Q$-function yields a lower value of $S$, even when the shape of the incoherent filter matches that of the target mode. These results demonstrate that the QPG enables coherent and programmable filtering of a broad class of structured temporal modes.

The minor deviation observed between the theoretical and reconstructed purities can be attributed to the combination of several experimental factors. In particular, for QPG filtering, the purity does not reach unity because the finite phase-matching bandwidth of the QPG limits its operational fidelity and therefore limits the achievable spectral purity of the filtered state. This effect can be mitigated by employing narrowband spectral filtering at the output of the QPG filtering channel \cite{santandrea2019njp}. In addition, slow temporal drifts during the long $Q$-function acquisition times, imperfect pump shaping, finite measurement resolution, and limited temporal and spectral windows can further contribute to the small mismatches observed for both QPG and incoherent filtering.

%The minor deviation between the theoretical coherent filter $Q$-function and the measured QPG filter $Q$-function. This can be attributed to the following reasons: imperfect shaping of pump QPG pump pulse, the finite phase-matching width of QPG reduces the fidelity of projecting or selecting into the selected mode, and temporal drifts over the long measurement time.

\section{Conclusion and Discussion}
In conclusion, we have experimentally demonstrated coherent temporal filtering of a spectrally multimode type-0 PDC source using a QPG and directly compared its performance with an analogous incoherent filtering scheme. Our results show that coherent filtering consistently extracts single well-defined Gaussian TMs with spectral purity exceeding 0.9 over a broad range of filter bandwidths. In contrast, incoherent filtering results in mixed TM states and the corresponding spectral purity decreases as the filter bandwidth increases. We have further showcased the versatility of QPG-based coherent filtering by extracting structured TMs, including HG modes and picosecond time-bin superposition states. These results show that QPG filtering enables the extraction of well-defined TMs from multimode PDC sources, which cannot be achieved using incoherent filtering.

The mode-selective filtering capability of QPG can enable high-visibility Hong-Ou-Mandel interference between photons generated by independent multimode sources, a key building block for quantum-network architectures. This capability can also enable the remote shaping of heralded single photons, which is difficult to achieve using conventional approaches such as PDC source engineering \cite{mosley2008prl, harder2013optexp, morrison2022aplphot, faleo2025arxiv} or spectral intensity filtering. This remote-shaping capability allows TM engineering at wavelengths where active pulse shaping of single photons is technically challenging. Furthermore, the QPG provides a route to selectively filter and read out TM-encoded information in the presence of noise, a functionality that cannot be achieved with phase-insensitive incoherent filtering \cite{thomas2026praapp}. Therefore, the demonstration of the QPG as a coherent temporal filter has significant implications for quantum communication, quantum sensing, and photonic quantum information processing systems.

\bmsection{Funding}
We acknowledge the European Union’s Horizon Europe research and innovation programme under grant agreement No 899587 (STORMYTUNE) for the funding.

\bmsection{Acknowledgment}
We thank Dr. Laura Ares Santos for helpful discussions. 

\bmsection{Data Availability}
All data required for evaluating our conclusions are present in the paper and supplementary material. Additional data related to this paper may be requested from the corresponding author.

\bmsection{Disclosures}
The authors declare no conflicts of interest

\bmsection{Supplemental document}
See Supplementary Material for supporting content.

%%%%%%%%%% If using BibTeX:
%\bibliography{sample}
\bibliography{ref}
%%%%%%%%%% If preparing manually:
% \begin{thebibliography}{1}
% \newcommand{\enquote}[1]{``#1''}

% \bibitem{Zhang:14}
% Y.~Zhang, S.~Qiao, L.~Sun, Q.~W. Shi, W.~Huang, L.~Li, and Z.~Yang,
%   \enquote{Photoinduced active terahertz metamaterials with nanostructured
%   vanadium dioxide film deposited by sol-gel method,}
%   {\protect\JournalTitle{Optics Express}} \textbf{22}, 11070--11078 (2014).

% \bibitem{Optica}
% {Optica}, \enquote{{Optica Publishing Group},}
%   \url{http://www.opg.optica.org}.

% \bibitem{FORSTER2007}
% P.~Forster, V.~Ramaswamy, P.~Artaxo, T.~Bernsten, R.~Betts, D.~Fahey,
%   J.~Haywood, J.~Lean, D.~Lowe, G.~Myhre, J.~Nganga, R.~Prinn, G.~Raga,
%   M.~Schulz, and R.~V. Dorland, \enquote{Changes in atmospheric consituents and
%   in radiative forcing,} in \enquote{Climate Change 2007: The Physical Science
%   Basis. Contribution of Working Group 1 to the Fourth Assesment Report of
%   Intergovernmental Panel on Climate Change,}  S.~Solomon, D.~Qin, M.~Manning,
%   Z.~Chen, M.~Marquis, K.~B. Averyt, M.~Tignor, and H.~L. Miler, eds.
%   (Cambridge University Press, 2007).

% \end{thebibliography}

\end{document}

\maketitle

\section{Joint Spectral Amplitude}
\begin{figure*}[b!]
	\centering
	\includegraphics[scale=1]{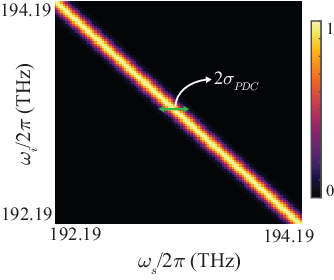}
	\caption{Simulated JSA of the PDC source. Here, the spectral correlation bandwidth $\sigma_{PDC}/{2\pi}$ is 0.04 THz.}\label{fig0}
\end{figure*}

The joint state of the signal and idler photons can be written as \cite{grice1997pra} %
\begin{equation}
	\ket{\psi}  = \int F(\omega_s,\omega_i) \ket{\omega_s}_s\ket{\omega_i}_i d\omega_sd\omega_i,\label{psi}
\end{equation}
where $F(\omega_s,\omega_i)$ is the joint spectral amplitude (JSA), and $\omega_s$ and $\omega_i$ are the angular frequencies of signal and idler photons, respectively. The JSA contains the complete time-frequency properties of the bi-photon state. For our type-0 PDC, the shape of the JSA is shown in Fig.~\ref{fig0}. Here, $\sigma_{PDC}$ represents the spectral correlation bandwidth.  

\section{Spectral density matrix of the heralded signal photon}
Here, we apply either coherent or incoherent filtering to the idler photon, which is used for heralding. We analyse the influence of filtering by evaluating the spectral density matrix of the heralded signal photon. 
\begin{figure*}[t!]
	\centering
	\includegraphics[width=\textwidth]{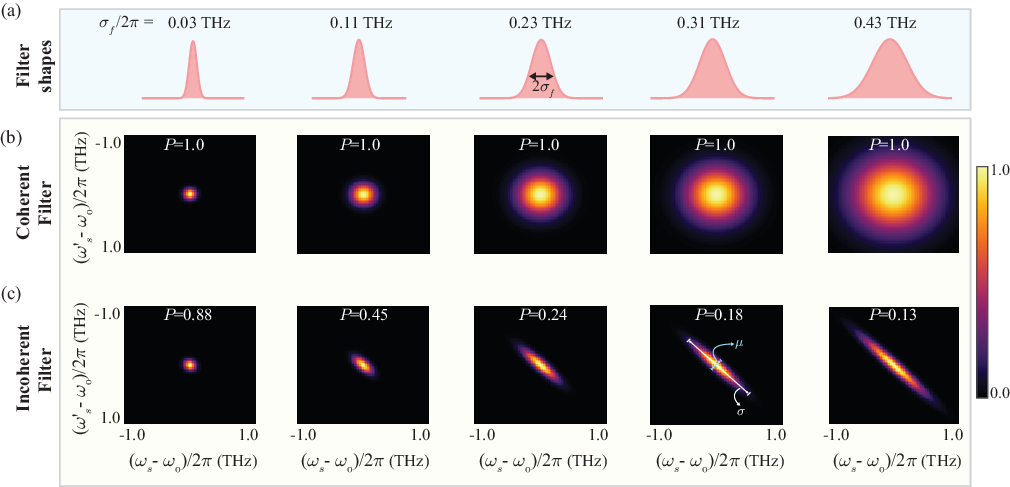}
	\caption{(a) Filter shapes for different bandwidths. (b) and (c) Spectral density matrix $\Gamma(\omega_s,\omega'_s)$ of the heralded signal photon corresponding to coherent and incoherent filtering, respectively. Here, $\sigma$ and $\mu$ are the spectral and coherence widths respectively. $\omega_0$ is the central frequency of the signal photon.}\label{fig0_1}
\end{figure*} 
\subsection{Coherent filtering}
We apply filtering to the idler photon and describe it by the operator $\hat{T}^{\mathrm{coh}}_{i}$ with a mode function $G(\omega_i)$. After filtering, we use the idler for heralding, and the corresponding heralded density matrix of the signal photon is given by  
\begin{equation}
	\hat{\rho}^{coh}_{s} = \mathrm{Tr}_i \left[ 
	\left( \hat{I}_s \otimes \hat{T}_i^{{coh}} \right)
	|\psi\rangle \langle \psi|
	\left( \hat{I}_s \otimes \hat{T}_i^{{coh}\dagger} \right)
	\right],\label{DM_coh}
\end{equation}
where $\hat{I}_s$ is the identity operator acting on the signal photon. The filter operator is given by
\begin{equation}
	\hat{T}^{coh}_{i} = \int G(\omega_i)G^{*}(\omega'_i) \iket{\omega_i}\ibra{\omega'_i}d\omega_id\omega'_i.\label{filter1}
\end{equation}
We substitute $\hat{T}^{coh}_{i}$ and $F(\omega_s,\omega_i)$ into Eq.~(\ref{DM_coh}), the heralded density matrix takes the form
\begin{equation}
	\hat{\rho}^{coh}_{s} = \iint \Gamma(\omega_s,\omega'_s) \sket{\omega_s}\sbra{\omega'_s}d\omega_sd\omega'_s,\label{DM1}
\end{equation}
where, $\Gamma(\omega_s,\omega'_s)$ denotes the spectral density matrix and is given by
\begin{equation}
	\Gamma(\omega_s,\omega'_s) = \iint F(\omega_s,\omega_i)F^*(\omega'_s,\omega'_i)G(\omega_i)G^{*}(\omega'_i) d\omega_id\omega'_i = C(\omega_s)C^*(\omega'_s).\label{DM2}
\end{equation}
We note that the spectral density matrix $\Gamma(\omega_s,\omega'_s)$ becomes a product of complex spectral amplitudes $C(\omega_s)C^*(\omega'_s)$, which implies that coherent filtering prepares the signal photon in a well-defined temporal mode (TM) \cite{brechtprx2015} $C(\omega_s)$ or in a coherent superposition of TMs. We find $\Gamma(\omega_s,\omega'_s)$ for a Gaussian shaped filter having different bandwidths as illustrated in Figure~\ref{fig1}(a). Here we vary the bandwidth of the filter over a broad range from the spectral correlation bandwidth of the PDC to nearly an order of magnitude larger. Figure~\ref{fig1}(b) shows the corresponding $\Gamma(\omega_s,\omega'_s)$. We find that for each $\Gamma(\omega_s,\omega'_s)$, the spectral width $(\sigma)$ and coherence width $(\mu)$ are equal, which is a signature of a single TM. Therefore, coherent filtering consistently yields a well-defined TM.  

\subsection{Incoherent filtering}
We apply an incoherent filter to the idler photon and use it for heralding. The filter operator $\hat{T}^{\mathrm{inc}}_{i}$ is given by
\begin{equation}
	\hat{T}^{inc}_{i} = \int G(\omega_i)G^{*}(\omega'_i) \delta(\omega_i-\omega'_i) \iket{\omega_i}\ibra{\omega'_i}d\omega_id\omega'_i.\label{filter2}
\end{equation}
The heralded density matrix of the signal photon takes the form
\begin{equation}
	\hat{\rho}^{inc}_{s} = \iint \Gamma(\omega_s,\omega'_s) \sket{\omega_s}\sbra{\omega'_s}d\omega_sd\omega'_s,\label{DM3}
\end{equation}
where, the spectral density matrix $\Gamma(\omega_s,\omega'_s)$ is given by
\begin{equation}
	\Gamma(\omega_s,\omega'_s) = \int F(\omega_s,\omega_i)F^*(\omega'_s,\omega_i)|G(\omega_i)|^2 d\omega_i \neq C(\omega_s)C^*(\omega'_s).\label{DM4}
\end{equation}
\begin{figure*}[t!]
	\centering
	\includegraphics[width=\textwidth]{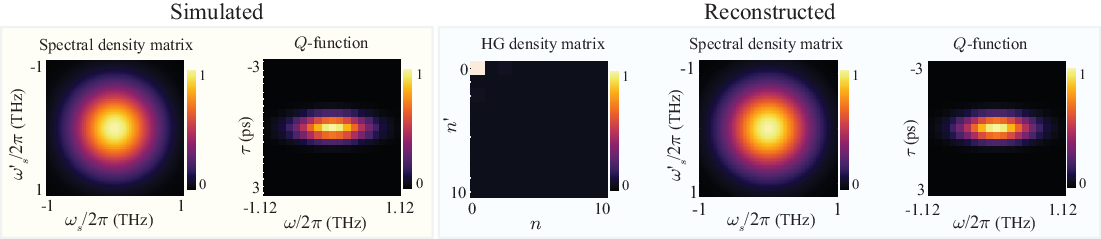}
	\caption{(a) Simulated spectral density matrix and the corresponding $Q$-function. (b) Reconstructed density matrix in the HG basis and spectral basis and the $Q$-function obtained from the reconstructed density matrix.}\label{fig1}
\end{figure*} 
We note that the spectral density matrix $\Gamma(\omega_s,\omega'_s)$ cannot be factorized into a product of complex spectral amplitudes $C(\omega_s)C^*(\omega_s)$, which is a signature of a multimode mixture of TMs, that is, a mixed state. Figure~\ref{fig1}(c) shows $\Gamma(\omega_s,\omega'_s)$ for different filter bandwidths. We find that when the filter bandwidth $\sigma_f$ is smaller than the PDC correlation bandwidth $\sigma_{PDC}$, the spectral and coherence widths remain approximately equal, which indicates an effectively single-mode state. As the filter bandwidth $\sigma_f$ increases the ratio of coherence width to spectral width decreases, which implies an increase in the degree of mixedness, that is, decrease in the spectral purity. This result has also been reported in the previous works \cite{christ2014pra, meyer2017pra}.

\subsection{Spectral purity}
The spectral purity of the heralded signal photon is given by
\begin{equation}
	P = \mathrm{Tr}\!\left[\left(\hat{\rho}^{{coh/inc}}_{s}\right)^2\right]= \iint |\Gamma(\omega_s,\omega'_s)|^2 d\omega_s d\omega'_s.\label{pur}
\end{equation}
Here, $P=1$ corresponds to a pure state, representing a well-defined temporal mode, whereas $P<1$ indicates a mixed state. The evaluated spectral purities for coherent and incoherent filtering are shown in Fig.~\ref{fig0_1}(b,c). Coherent filtering consistently yields high-purity single-mode states, whereas for incoherent filtering the purity decreases with increasing filter bandwidth.

\section{Chronocyclic $Q$-function}
The chronocyclic $Q$-function is defined as the projection of the input TM described by spectral density matrix $\Gamma(\omega_s,\omega'_s)$ onto the analog of a coherent state in TF phase space, that is, a Fourier-limited Gaussian mode $\mathcal{E}_c(\omega_s,\omega,\tau;\sigma_c)$ of bandwidth $\sigma_f$ with spectral shift $\omega$ and time delay $\tau$. To emulate a quadrature phase space $Q$-function, where the axes are dimensionless and the coherent state has equal width along both axes, we apply the rescaling  
\begin{equation*}
	\Omega = \frac{\omega}{\sigma_f}, \quad T = \tau \sigma_f, \quad \Omega_s = \frac{\omega_s}{\sigma_f}.
\end{equation*}
We now express the chronocyclic $Q$-function $Q(\Omega,T)$ in terms of spectral density matrix $\Gamma(\Omega_s,\Omega'_s)$ as \cite{bhattacharjee2025optexp}
\begin{equation}
	Q(\Omega,T) = \iint \Gamma(\Omega_s,\Omega'_s)\mathcal{E}^*_c(\Omega_s,\Omega,T)\mathcal{E}_c(\Omega'_s,\Omega,T)d\Omega_sd\Omega'_s,\label{Qfun2}
\end{equation} 
where, 
\begin{equation}
	\mathcal{E}_c(\Omega'_s,\Omega,T) = \exp\left[-\frac{(\Omega_s-\Omega_0-\Omega)^2}{2}+i(\Omega_s-\Omega_0)T\right].\label{coh-TF}
\end{equation} 
Here, $\Omega_0=\frac{\omega_0}{\sigma_f}$ and $\omega_0$ is the central angular frequency. We note that the chronocyclic $Q$-function contains complete information about the spectral density matrix. In the experiment, we measure the chronocyclic $Q$-function of the heralded signal photon and retrieve the spectral density matrix by employing the following procedure.

\section{Relation between spectral purity and bandwidth product of the chronocyclic $Q$-function}
In this section, we derive an analytical relation between the spectral purity $P$ and the bandwidth product $\Delta\Omega\times\Delta T$ of the marginal distributions of the chronocyclic $Q$-function. This relation provides additional insight into the physical significance of the bandwidth product discussed in the main text but is not used for the purity estimation presented in the manuscript. We consider partially coherent TMs described by
\begin{equation}
	\Gamma(\omega_s,\omega'_s)=
	N\exp\left[-\frac{(\omega_s-\omega_0)^2+(\omega'_s-\omega_0)^2}{2\sigma_s^2}\right]
	\exp\left[-\frac{(\omega_s-\omega'_s)^2}{2\sigma_c^2}\right],
	\label{DM5}
\end{equation}
where $\sigma_s$ and $\sigma_c$ denote the spectral bandwidth and coherence bandwidth, respectively, and $N$ is the normalization constant. Equation~(\ref{DM5}) models the partially coherent TMs generated by incoherent Gaussian filtering. In the limit $\sigma_c\rightarrow\infty$, the spectral density matrix reduces to that of a Fourier-limited Gaussian TM.

The spectral purity of $\Gamma(\omega_s,\omega'_s)$, evaluated using Eq.~(\ref{pur}), is given by

\begin{equation}
	P=\frac{1}{\sqrt{1+\frac{2\sigma_s^2}{\sigma_c^2}}}.
	\label{purity}
\end{equation}
For a Fourier-limited Gaussian TM, the spectral purity reaches unity $P=1$. As the ratio $\sigma_s/\sigma_c$ increases, the spectral purity decreases.

Substituting Eq.~(\ref{DM5}) into Eq.~(\ref{Qfun2}) and choosing $\sigma_f=\sigma_s$, the chronocyclic $Q$-function becomes

\begin{equation}
	Q(\Omega,T)=
	A\exp\left[-\frac{(\Omega-\omega_0)^2}{2}\right]
	\exp\left[-\frac{T^2}{2\left(1+\frac{\sigma_s^2}{\sigma_c^2}\right)}\right].
	\label{Qfun3}
\end{equation}
The bandwidth product of the marginal distributions of the $Q$-function is
\begin{equation}
	\Delta\Omega\times\Delta T=
	\sqrt{1+\frac{\sigma_s^2}{\sigma_c^2}}.
	\label{Qfun4}
\end{equation}
In the fully coherent limit $(\sigma_c\rightarrow\infty)$, the bandwidth product reaches its minimum value $\Delta\Omega\times\Delta T=1$, corresponding to a Fourier-limited Gaussian TM. As $\sigma_s/\sigma_c$ increases, the coherence decreases and the bandwidth product increases. This behavior is reflected in the broadening of the $Q$-function. Combining Eqs.~(\ref{purity}) and (\ref{Qfun4}) yields

\begin{equation}
	P=\frac{1}{\sqrt{2(\Delta\Omega\times\Delta T)^2-1}}.
	\label{Qfun5}
\end{equation}
Equation~(\ref{Qfun5}) shows that the spectral purity decreases monotonically with increasing bandwidth product.

\section{Density Matrix Reconstruction Procedure}\label{sec2}
The density operator of the heralded signal photon, denoted $\hat{\rho}$, can be represented in different bases. In the frequency basis, its matrix elements are given by the spectral density matrix $\Gamma(\omega_s,\omega'_s) = \langle \omega_s | \hat{\rho} | \omega'_s \rangle$. Here, we  reconstruct $\rho$ in the Hermite-Gaussian (HG) mode basis $\{ |u_n\rangle \}_{n=0}^{d-1}$, where the density matrix has elements $\rho_{mn} = \langle u_m | \hat{\rho} | u_n \rangle$. The representations in these two bases are related by
\begin{equation}
	\Gamma(\omega_s,\omega'_s) = \sum_{m,n} \rho_{mn}\, u_m(\omega_s)\, u_n^*(\omega'_s),\label{hg1}
\end{equation}
where, 
\begin{equation*}
	u_n(\omega_s) = \frac{1}{\sqrt{2^n n!\sqrt{\pi}\sigma_s}}H_n\left(\frac{\omega_s-\omega_0}{\sigma_s}\right)e^{-\frac{(\omega_s-\omega_0)^2}{2\sigma_s^2}}.\label{hg}
\end{equation*}
The reconstruction procedure determines the matrix elements $\rho_{mn}$ from measured chronocyclic $Q$-function data in the following manner. The chronocyclic $Q$-function sampled over a discrete set of TF phase-space points $(\Omega_k, T_k)$, where each index $k$ represents a projection of the heralded signal photon onto a known TF coherent probe mode $\mathcal{E}_c(\Omega_s,\Omega_k,T_k)$ defined by a frequency shift $\Omega_k$ and a temporal delay $T_k$. This process is formally described by a positive operator-valued measure (POVM) element $\Pi_k$, such that the measured value of the chronocyclic $Q$-function is given by
\begin{equation}
	Q_k \equiv Q(\Omega_k,T_k) = \mathrm{Tr}(\Pi_k \rho).
\end{equation}
The density matrix is estimated by minimizing the weighted squared deviation between the measured normalized $Q$-function values $\tilde{Q}_k$ and the model predictions,
\begin{equation}
	\rho = \arg\min_{\rho} \sum_k w_k \left[\mathrm{Tr}(\Pi_k \rho) - \tilde{Q}_k \right]^2.
\end{equation}
In this minimization, we impose the physical constraints that $\rho$ is positive semidefinite and has unit trace. The weighting factors $w_k$ are chosen to approximate Poissonian counting statistics, effectively reducing the influence of data points with higher relative variance. The matrix elements $\rho_{mn}$ in the HG basis are obtained directly from this optimization procedure. We retrieve the spectral density matrix using Eq.~(\ref{hg1}).

\section{Reconstruction with simulated theoretical $Q-$function data}
To assess the performance of the reconstruction protocol, we first analyze simulated chronocyclic $Q$-function data. We choose simulation parameters, including resolution and the time-frequency (TF) phase space window are comparable to those employed in the experiment. Using the simulated data, we can identify reconstruction artifacts that arise solely from the finite measurement window and sampling resolution.

\begin{figure*}[t!]
	\centering
	\includegraphics[width=\textwidth]{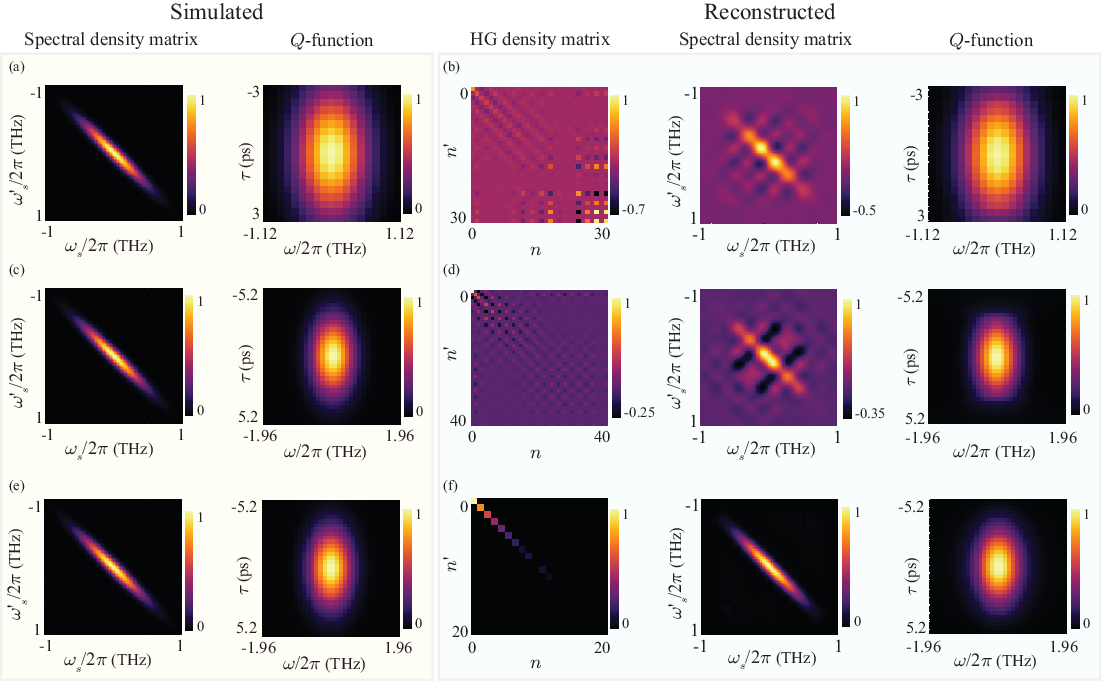}
	\caption{(a) Simulated spectral density matrix and the corresponding $Q$-function. (b) Reconstructed density matrix in the HG basis and spectral basis and the $Q$-function obtained from the reconstructed density matrix. (c) Simulated spectral density matrix and the corresponding $Q$-function. (d) Reconstructed density matrix in the HG basis and spectral basis and the $Q$-function obtained from the reconstructed density matrix. (e) Simulated spectral density matrix and the corresponding $Q$-function. (f) Reconstructed density matrix in the HG basis and spectral basis and the $Q$-function obtained from the reconstructed density matrix.}\label{fig2}
\end{figure*} 
\subsection{For coherent filtering}
Figure~\ref{fig1}(a) shows the theoretical spectral density matrix and the corresponding $Q$-function for a filter bandwidth $\sigma_f/{2\pi} = 0.41$ THz. We apply the maximum-likelihood  procedure on the simulated $Q$-function. Figure~\ref{fig1}(b) shows the reconstructed density matrix expressed in the HG basis and in the frequency basis. We find excellent agreement between the reconstructed and theoretical density matrices. Furthermore, the $Q$-function obtained from the reconstructed state is nearly identical to the theoretical profile, yielding a similarity of $S = 1$. These results confirm that with the given experimental resolution and sampling window, the reconstruction procedure reliably retrieves the expected single-mode state.
\begin{figure*}[t!]
	\centering
	\includegraphics[width=\textwidth]{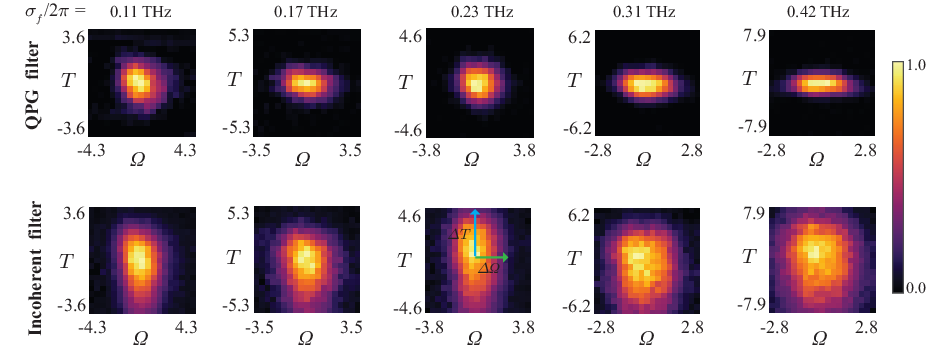}
	\caption{Measured $Q-$function data corresponding to QPG and incoherent filtering for different filter bandwidth $\sigma_f$.}\label{fig3}
\end{figure*} 
\begin{figure*}[t!]
	\centering
	\includegraphics[width=\textwidth]{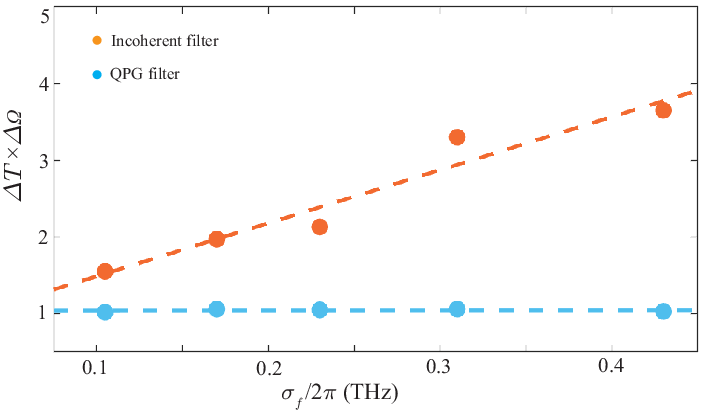}
	\caption{Bandwidth product $\Delta \Omega \times \Delta T$ extracted from the measured chronocyclic $Q$-functions as a function of the filter bandwidth $\sigma_f$/2$\pi$ for QPG filtering (blue) and incoherent filtering (orange). Dashed lines are guides to the eye.
	}\label{fig5}
\end{figure*} 
\begin{figure*}[t!]
	\centering
	\includegraphics[width=\textwidth]{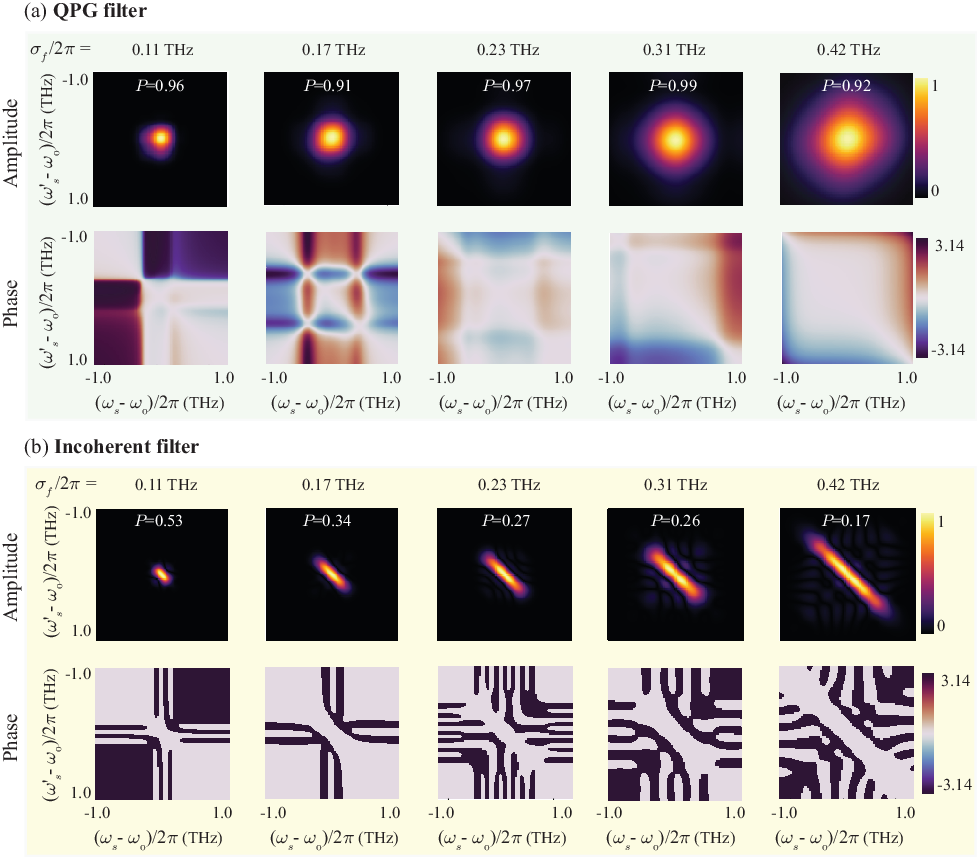}
	\caption{(a) and (b) Show the reconstructed spectral density matrix corresponding to QPG and incoherent filtering for different filter bandwidth $\sigma_f$.}\label{fig4}
\end{figure*} 
\subsection{For incoherent filtering}
Figure~\ref{fig2}(a) shows the theoretical spectral density matrix and the corresponding $Q$-function for a filter bandwidth $\sigma_f/{2\pi} = 0.41$ THz. Within the experimental phase-space sampling window, the simulated $Q$-function is partially truncated. Figure~\ref{fig2}(b) shows the reconstructed density matrix in the HG basis and in the frequency basis. We find spurious contributions from higher-order HG modes, which yield structural features in the reconstructed density matrix. Nevertheless, the reconstructed $Q$-function shows excellent agreement with the theoretical profile, with a similarity of $S = 0.99$, which indicates that the simulated $Q$-function is reproduced with fidelity from the reconstructed density matrix despite the presence of structural features. %In this reconstruction, the HG basis dimensionality $d$ is incremented until the similarity between the reconstructed and theoretical $Q$-functions reaches a threshold of $S \geq 0.98$ and at that $d$ the basis dimensionality is fixed. 
The observed structural artifacts arise from the limited phase-space window, which introduces a non-uniqueness in the reconstruction procedure. Specifically, over a restricted sampling region, different HG modes can produce nearly identical responses or $Q$-function profiles, which prevents the maximum-likelihood estimator from uniquely distinguishing between them.

To resolve this issue, we repeat the reconstruction using simulated $Q$-function data in an extended phase-space window, as shown in Fig.~\ref{fig2}(c). The reconstructed density matrix in Fig.~\ref{fig2}(d) no longer exhibits a dominant contribution from higher-order HG modes, confirming that the main artifact originates from the limited phase-space window. However, weak residual structure remains even for the extended window, while the reconstructed $Q$-function still shows excellent agreement with the theoretical profile. Since these results are obtained from ideal simulated data, the remaining features can be attributed to numerical reconstruction effects associated with finite window size and finite sampling resolution. 

To suppress these numerical artifacts and stabilize the estimation of experimental data, we introduce a restricted reconstruction model where the density matrix is constrained to be diagonal in the HG basis. This assumption is physically justified by the fact that HG modes closely approximate the Schmidt modes of the JSA in Fig.~\ref{fig0}. Imposing this constraint also reduces the number of free parameters in the reconstruction procedure. Figure~\ref{fig2}(f) shows the reconstructed density matrix and the corresponding $Q$-function obtained using this restricted model. We find excellent agreement between the reconstructed and simulated spectral density matrices, and the reconstructed $Q$-function shows a perfect match with the theoretical profile ($S = 1$).

We adopt this restricted reconstruction approach only for the subsequent analysis of the incoherent-filtering data. A fully unconstrained reconstruction can introduce numerical instabilities in the presence of highly multimode mixtures, and addressing these issues lies outside the primary scope of this study. By using the restricted model, we obtain stable and consistent estimates of the spectral purity, which is the central metric of this work.

\section{Measured $Q-$function data and reconstructed spectral density function}
Figure~\ref{fig3} shows the measured Q-functions for QPG and incoherent filtering corresponding to different filter bandwidths. In the main manuscript Fig.~3(b,c), we present the same data using equal spans along the $\Omega$ and $T$ axes in order to highlight the symmetry properties of the $Q$-function. We note that $\Delta T$ and $\Delta\Omega$ are representing the bandwidths of the marginal axes $T$ and $\Omega$ respectively. We further evaluate the bandwidth product $\Delta\Omega \times \Delta T$ as a function of the filter bandwidth $\sigma_f$ as shown in Fig.~\ref{fig5}. For QPG filtering, $\Delta\Omega \times \Delta T$ remains nearly constant across the investigated bandwidth range and stays close to unity, consistent with a Fourier-limited Gaussian TM. In contrast, for incoherent filtering, $\Delta\Omega \times \Delta T$ increases with $\sigma_f$, reflecting the increasing extent of the $Q$-function in TF phase space and indicating the presence of multimode mixtures. Here, we present the complete data set for all measured Q-functions and use them for reconstructing the spectral density matrix. We note that, for incoherent filtering, the $Q$-functions are partially truncated within the sampled phase-space window. As discussed in Sec.~\ref{sec2}, such truncation can introduce artifacts in the reconstructed spectral density matrix. To mitigate this effect, a constant background level estimated from the corner regions of each measured Q-function is first subtracted. The Q-function is then extended beyond the measured phase-space window using a quadratic-log tail extrapolation obtained by fitting the outer region of the measured distribution. The extrapolated distribution is used only outside the measured window, while the experimentally measured data within the original window remain unchanged prior to reconstruction.

Figure ~\ref{fig4}(a) and ~\ref{fig4}(b) show the reconstructed spectral density matrices obtained from the measured Q-function data using the maximum-likelihood estimation method described in Sec.~\ref{sec2}. For QPG filtering, the reconstructed density matrices exhibit nearly equal spectral and coherence widths, which indicates a single TM state. The spectral purity remains above 0.9 across the sampled bandwidth range, confirming that QPG filtering yields high-purity TMs. On the other hand, for incoherent filtering, increasing the filter bandwidth leads to a progressive reduction in the ratio of coherence width to spectral width and a corresponding decrease in spectral purity.

% Bibliography
\bibliography{ref_1}

%Manual citation list
%\begin{thebibliography}{1}
%\bibitem{Zhang:14}
%Y.~Zhang, S.~Qiao, L.~Sun, Q.~W. Shi, W.~Huang, %L.~Li, and Z.~Yang,
 % \enquote{Photoinduced active terahertz metamaterials with nanostructured
  %vanadium dioxide film deposited by sol-gel method,} Opt. Express \textbf{22},
  %11070--11078 (2014).
%\end{thebibliography}